\documentclass[a4paper,twocolumn]{Gaia2004}
\usepackage{times}
\usepackage{epsfig}
\usepackage{natbib}
\title{From Detailed Galaxy Simulations to a Realistic End-of-Mission Gaia
  Catalogue}
\author{A.G.A.~Brown}
\affil{Sterrewacht Leiden,  P.O.\ Box 9513, 2300 RA Leiden, The Netherlands}
\author{H.M.~Vel\'azquez}
\author{L.A.~Aguilar}
\affil{Instituto de Astronom{\'\i}a, UNAM, Apartado Postal 877, Ensenada, 22800,
Baja California, Mexico}
\begin{document}
\maketitle
\begin{abstract}
We address the problem of identifying remnants of satellite galaxies in the
halo of our galaxy with Gaia data. We make use of N-body simulations of dwarf
galaxies being disrupted in the halo of our galaxy combined with a Monte Carlo
model of the Milky Way galaxy. The models are converted to a simulated Gaia
catalogue containing a realistic number ($\sim10^8$--$10^9$) of stars. The
simulated catalogue can be used to study how to handle the large data set that
Gaia will provide and to study issues such as how to best retrieve information
on substructure in the Galactic halo. The techniques described are applicable
to any set of N-body simulations of (parts of) the Galaxy.
\end{abstract}
\section{Introduction}
Cosmological models predict that halos of luminous galaxies are assembled
through merging of smaller structures. Remnants of past mergers will survive
for a long time in the halo as debris streams because of the very long
dynamical time scale. The Gaia mission offers a unique opportunity to search
for and study these remnants with full phase space information in our galaxy's
halo. However disentangling the possibly many remnants from each other and
from the background of Galactic stars will be very challenging. The Gaia
catalogue will contain about one billion objects of which the stars in the
individual debris streams may form only a very small fraction. In addition the
streams are spread out all over the sky. Identifying them will require a
combination of search methods that make use both of conserved dynamical
quantities of the debris stream \citep[such as energy and angular momentum,
see e.g.,][]{Helmi1999,Helmi2000} and astrophysical properties of the
constituent stars (from photometric data).

The goals of this work are: \textbf{1.} study the retrieval from Gaia data of
remnants of satellite galaxies that have been disrupted in the potential of
our Galaxy; \textbf{2.} include a realistic model of the Galactic background
population against which the remnants have to be detected; \textbf{3.} gain
practical experience with analysing and visualising the enormous volume of
information that will be present in the Gaia data-base.

To achieve this we built a Monte Carlo model of the smooth components of the
Galaxy containing a realistic number of stars and performed tree-code $N$-body
simulations of satellites that are disrupted while orbiting the Galaxy. The
Galaxy and satellite models were subsequently combined and Gaia observations
were simulated. In the following we summarise how the simulated Gaia survey
was generated, with emphasis on the proper combination of the satellite and
Milky Way models, and we present some examples of the simulated Gaia data. For
details we refer to \cite{Brown2004}.

\section{Monte Carlo Model of our Galaxy}

The Milky Way model consists of three spatial components: a bulge with a
Plummer density law, a double exponential disk, and a flattened halo
($c/a=0.8$) for which the density drops as $r^{-3.5}$. The kinematics are
modelled in a very simple manner: each component rotates with constant
velocity dispersion, for the bulge an isotropic dispersion is assumed, for the
disk a different velocity ellipsoid is used for each spectral type (OBAFGKM),
while for the halo the dispersions are the same for each spectral type. The
individual stars are assigned an absolute magnitude $M_V$ and a colour $(V-I)$
from a Hess-diagram which is considered fixed for all Galactic components and
all positions throughout the Galaxy. The Hess-diagram is taken from Table 4--7
of \citet{MB81} and provides the relative numbers of stars in bins of absolute
magnitude ($M_V$) and spectral type. These numbers integrated over spectral
type provide the luminosity function. Although this is a highly simplified and
certainly not self-consistent model of the Galaxy it is good enough for
providing the `background' distribution in phase space against which the
debris streams have to be found.

\begin{figure}
\epsfig{file=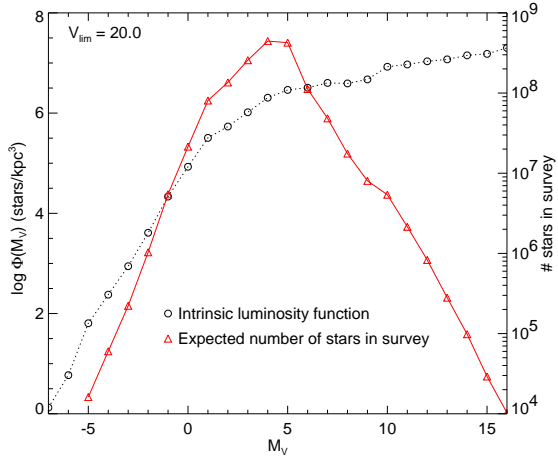,width=\columnwidth}
\caption{Intrinsic (circles, left vertical scale) and weighted (triangles,
  right vertical scale) luminosity functions for our Milky Way model plotted
  vs.\ $M_V$. The weighted luminosity function is plotted in terms of the
  number of stars expected in the simulated Gaia survey.}
\label{fig:lumfunc}
\end{figure}

For a magnitude limited survey, as is the case for Gaia, a straightforward
Monte Carlo realisation of the Galactic model is potentially a very wasteful
procedure. Most simulated stars will be too faint to be included in the
survey. Hence we used a strategy that minimizes wasted effort by generating
stars only within the magnitude limited sphere centred on the observer. This
results in a Galaxy model with a luminosity function that is weighted by the
space density integrated over the volume limited sphere for each spectral
type. This luminosity function is shown in Fig.~\ref{fig:lumfunc}.
Nevertheless the generation of the Monte Carlo model is still a large
computational task and in order to save time we decided not to simulate the
part of the sky within Galactic coordinates: $-90^\circ\leq\ell\leq +90^\circ$
and $-5^\circ\leq b\leq +5^\circ$. Tracing the debris streams in this part of
the sky will be difficult in practice due to the large extinction in those
directions. In our Galactic model $\sim80$\% of the stars lie in this region
of the sky as seen from the Sun for a survey limited at $V=20$. The resulting
simulated survey of the Galaxy is fully sampled and contains $3.2\times10^8$
stars. For each of these we generated the 6 phase space coordinates (positions
and velocities), an absolute magnitude, a population type (bulge, disk or
halo) and a spectral type.

\section{$\mathbf{N}$-body Models of Dwarf Galaxies}

\begin{figure}
\epsfig{file=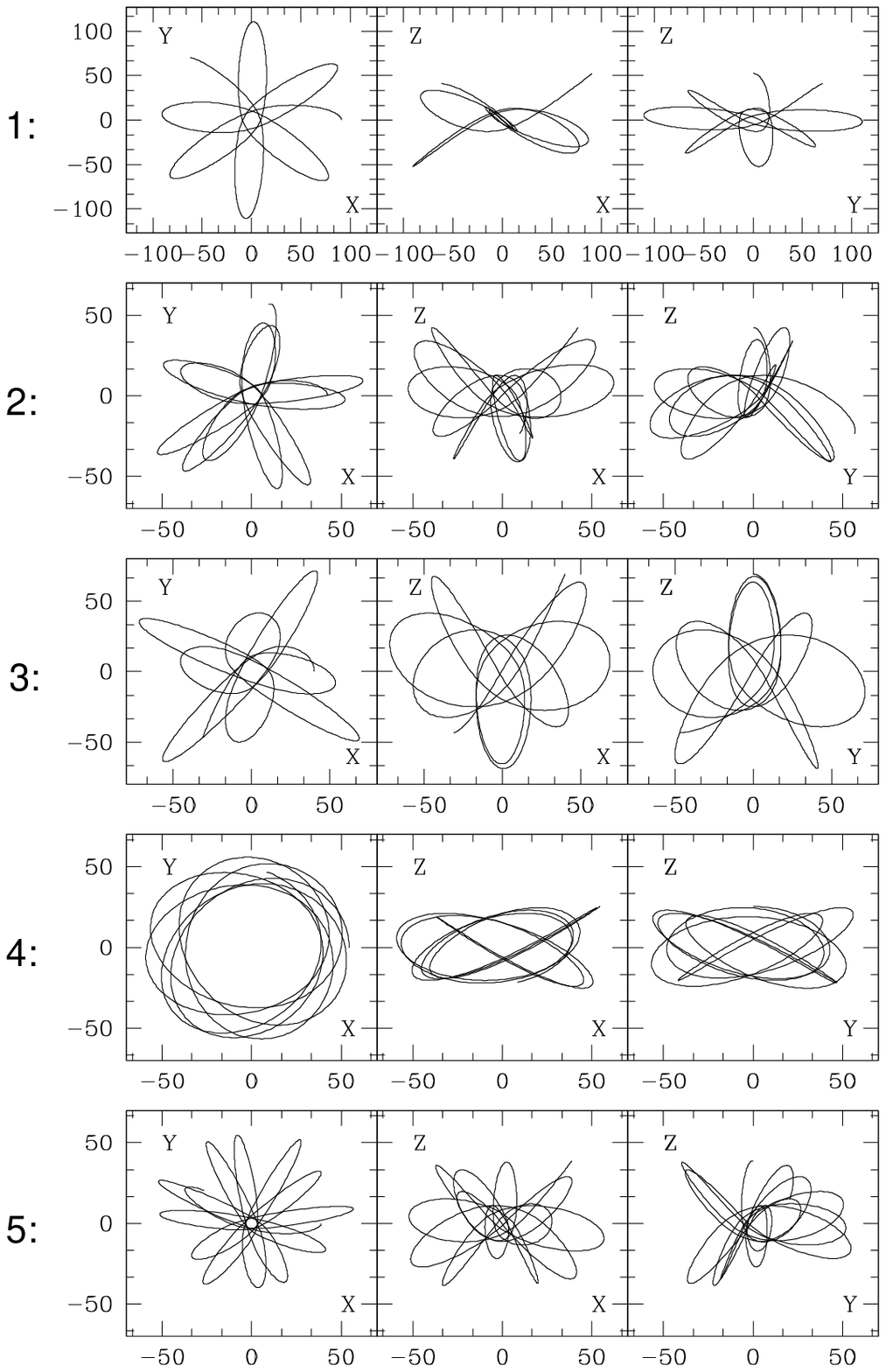,width=\columnwidth,clip=}
\epsfig{file=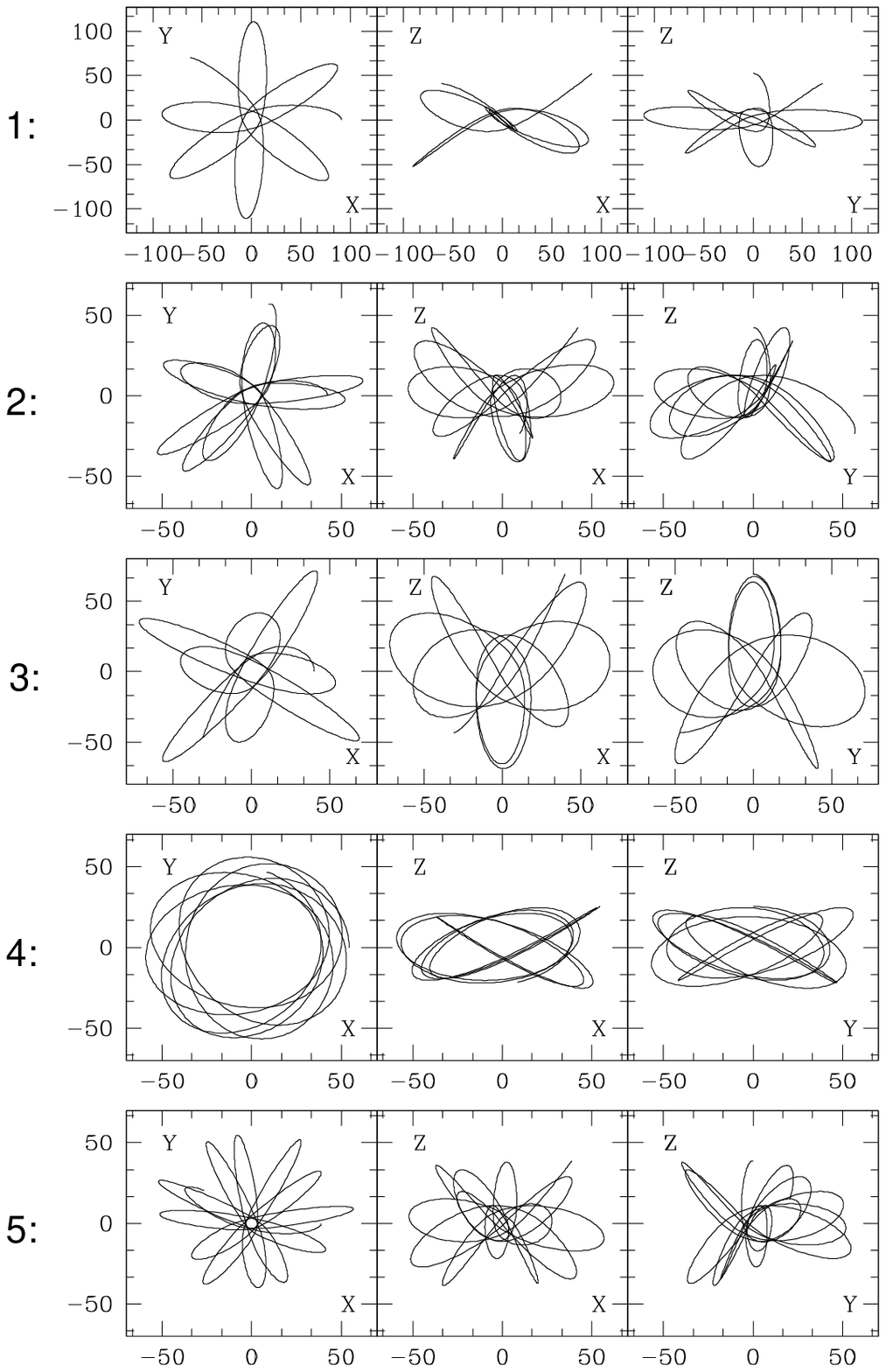,width=\columnwidth,clip=}
\caption{Satellite orbits for models no.\ 1 and 4 projected onto the principal
axes of our Milky Way galaxy model. The Galactic plane corresponds to the XY
projection. Distances are in kpc.}
\label{fig:satorbits}
\end{figure}

The models of the debris streams were generated by simulating the disruption
of dwarf galaxies orbiting our galaxy. The Milky Way is represented by a rigid
potential which is derived from a mass-model consisting of: a double
exponential disc, a spherical bulge with Hernquist profile, and a logarithmic
Halo potential with flattening $c/a=0.8$. The dwarf galaxies are represented
by King models \citep{King1966} with a mass of $5.6\times10^6$~M$_\odot$ or
$2.8\times10^6$~M$_\odot$ and $10^6$ particles. The tidal radius and
concentration parameter $c=\log(r_t/r_c)$ are 3150~pc and $0.9$,
respectively. The satellites are placed on five different orbits which vary in
apocentre, pericentre and the initial inclination with respect to the Milky
Way's disc. The simulations were evolved with a tree-code for
$\sim10$~Gyr. Multiple debris streams can be simulated by combining different
$N$-body snapshots (different orbits and/or ages), and each snapshot can be
rotated around the $Z$-axis or flipped with respect to the disc plane. Two of
the orbits for the $N$-body satellites are shown in
Fig.~\ref{fig:satorbits}. All stars of a given dwarf galaxy are assumed to be
of the same age. Masses are drawn from a mass-function and $M_V$ and $(V-I)$
are derived from a low-metallicity isochrone \citep[from][]{Girardi2000} of
the appropriate age.
\section{Modelling the Gaia Survey}
\begin{figure}
\begin{center}
\epsfig{file=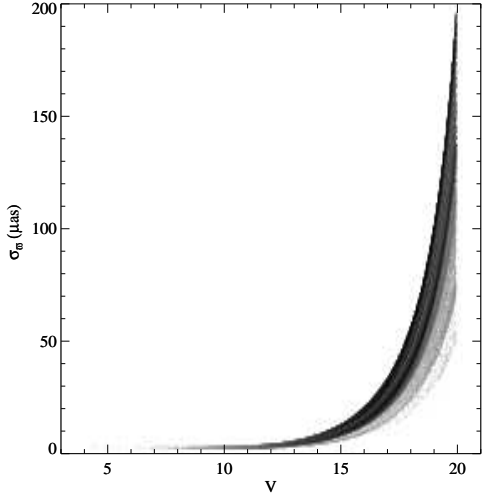,width=0.8\columnwidth}
\end{center}
\caption{Parallax errors vs.\ $V$-magnitude in the simulated Gaia catalogue
  for 1 million stars from the model Milky Way. The grey scale indicates the
  number of stars. The spread in the values at each $V$ is caused by the
  colour dependence of the errors (see Eq.(\ref{eq:parerr})).}
\label{fig:parerr}
\end{figure}

The 6 phase space coordinates $(\mathbf{r},\mathbf{v})$ for each star in the
Galaxy and satellite model are referred to the Solar position and velocity and
converted to the 5 astrometric parameters; position $(\ell,b)$, parallax
$\varpi$, proper motions $\mu_{\ell*}=\mu_\ell\cos b$ and $\mu_b$, and the
radial velocity $v_\mathrm{rad}$. The conversion is done using the standard
prescriptions for transforming Cartesian position and velocity coordinates
into astrometric parameters and radial velocities, such as described in in
Volume~1, Section~1.5.6, of the Hipparcos Catalogue \citep{ESA1997}. The
astrometric errors are added as a function of $G$ (Gaia broad band magnitude)
and $(V-I)$:
\begin{eqnarray}
\label{eq:parerr}
\sigma_\varpi \simeq &
(7+105z+1.3z^2+6\times10^{-10}z^6)^{1/2} \nonumber \\
 & \times[0.96+0.04(V-I)] \,,
\end{eqnarray}
where $z=10^{0.4(G-15)}$ and
\begin{eqnarray}
G = & V+0.51-0.50\times\sqrt{0.6+(V-I-0.6)^2}- \nonumber \\
    & 0.065\times(V-I-0.6)^2 \,.
\end{eqnarray}
The parallax errors for the simulated Gaia data of the Milky Way model are
shown in Fig.~\ref{fig:parerr}. The mean position and proper motion errors are
$\sigma_0=0.87\sigma_\varpi$ and $\sigma_\mu=0.75\sigma_\varpi$, respectively,
and the variation of the astrometric errors with ecliptic latitude $\beta$ is
included. The radial velocity errors for OBA-type stars are $0.25$, $4$, $10$,
$50$~km~s$^{-1}$ at $V=10$, $15$, $16$, $17$, respectively. For FGKM-type
stars the errors are $0.1$, $1$, $2$, $6$ km~s$^{-1}$ at $V=10$, $16$, $17$,
$18$. For stars fainter than $V=17$ (OBA) or $V=18$ (FGKM) radial velocities
are not available. These error prescriptions are according to the Gaia Concept
and Technology Study Report \citep{ESA2000}.

\subsection{Combining the Simulated Milky Way and Satellite data}

Having made a considerable effort to realistically simulate the number of
Galaxy stars that is expected to be seen by Gaia, we want to ensure that the
satellite simulations are properly added to the Galaxy data. This means that
the number of satellite particles in our simulated Gaia catalogue should be a
realistic fraction of the number of Galactic particles. Getting this right is
not trivial and we explain our solution to this problem here.

Given a certain distribution of stars along the orbit of a particular dwarf
galaxy (corresponding to an $N$-body `snap-shot'), the number of stars from
this satellite that will end up in the Gaia catalogue depends on three
factors: \textbf{1.} The overall number stars (i.e., luminous particles) in
the dwarf galaxy. This number is determined by its overall luminosity and the
stellar mass function. \textbf{2.} The Gaia survey limit leads to an upper
limit $M_{V,\mathrm{max}}$ on the absolute magnitude of visible satellite
stars. \textbf{3.} The variation of $M_{V,\mathrm{max}}$ along the satellite
orbit, caused by a variation in distance from the Sun.

\begin{figure}
\begin{center}
\epsfig{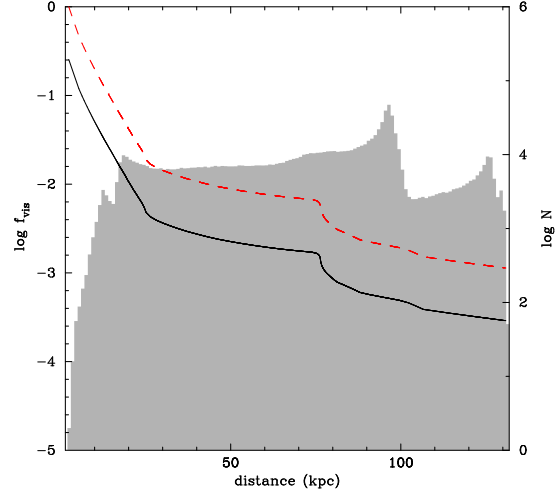}
\end{center}
\caption{$f_i$ vs $s_i$ for satellite model 1. The shaded histogram (right
 vertical scale) shows the distribution of $s_i$. The solid line (left
 vertical scale) shows $f_i(s_i)$. Features in this line reflect features in
 the luminosity function. The dashed line shows the $f_i(s_i)$ when assuming
 all $N$-body stars to be brighter than the faintest satellite star that
 enters the Gaia survey. We calculated $f_i$ using a 10 Gyr isochrone with
 $Z=0.004$ and $\xi(m)\propto m^{-1.5}$. The overall fraction of visible
 $N$-body particles is only $2.5\times10^{-3}$ (!) and $10^{-2}$ for the two
 cases considered.}
\label{fig:visfrac}
\end{figure}

At each distance $s_i$ where an $N$-body particle is located the fraction of
visible satellite stars $f_i$ can be calculated:
\begin{equation}
f_i(s_i)=\frac{\int_{m(M_{V,\mathrm{max}})}^{m_\mathrm{up}}\xi(m)dm}{\int
  \xi(m)dm}\,
\end{equation}
where $M_{V,\mathrm{max}}(s_i)=V_\mathrm{lim}-5\log s_i+5$, while $\xi(m)$ is
the mass function and $m_\mathrm{up}$ its upper limit. The overall fraction
$f$ of visible $N$-body stars is $f=\sum_i f_i(s_i)/N_\mathrm{sim}$. As
Fig.~\ref{fig:visfrac} illustrates the overall fraction of visible stars for a
fully populated mass function can be very small depending on the distribution
of satellite stars along the orbit.
In addition we are faced with the problem that real dwarf galaxies contain up
to $10^9$ stars but our $N$-body models contain 1 million particles only. Of
these `expensive' $N$-body particles we want to waste as little as
possible. An obvious step is to assume that all N-body particles represent
stars that are brighter than the faintest star that can enter the Gaia survey
given the distance distribution of the $N$-body particles. This will raise the
overall visible fraction $N$-body particles as illustrated by the dashed line
in Fig.~\ref{fig:visfrac}.  

This hints at the following solution: assume that all $N$-body particles
represent a bright tracer population, such as AGB stars. That is, $M_V\leq
M_{V,\mathrm{tracer}}$ for all particles. This will further raise the overall
visible fraction $f$. The details of this procedure are discussed in
\cite{Brown2004} where it is shown that depending on the total luminosity
(mass) of the simulated satellite one can then easily retain the majority of
the $N$-body particles while still preserving the variation of the visible
fraction of tracer stars along the debris stream.

\section{Tracing Debris Streams in the Gaia Catalogue}

\begin{figure*}[t]
\begin{center}
\epsfig{file=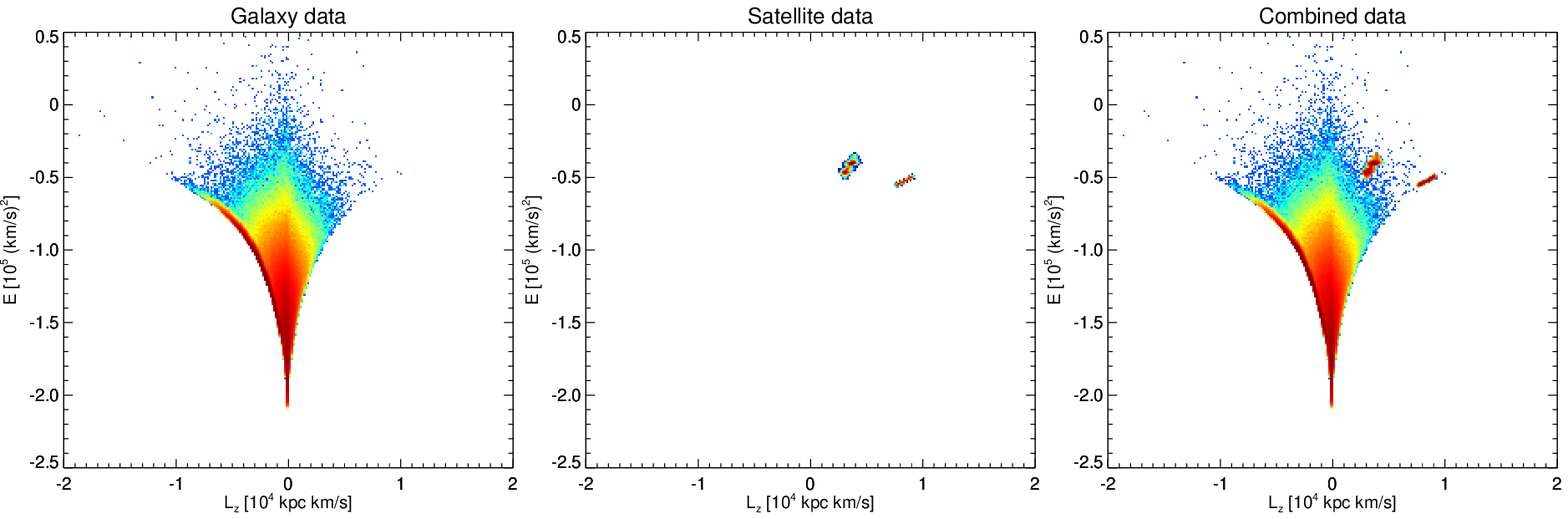,width=0.9\textwidth}
\epsfig{file=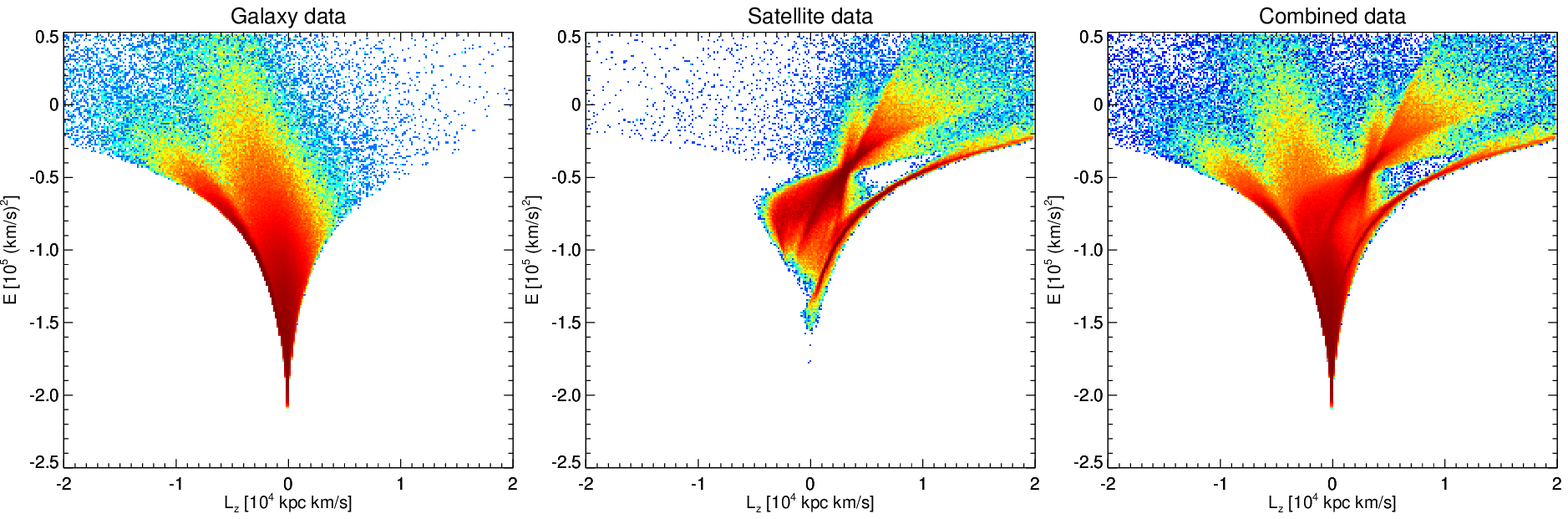,width=0.9\textwidth}
\end{center}
\caption{Energy vs angular momentum diagrams. The top row shows the true
  (simulated) $E$-$L_z$ distribution and the bottom row the energy and angular
  momentum as they would would appear when derived from the Gaia
  catalogue. From left to right the panels show the data for the Monte Carlo
  model of the Milky, for the two satellites (models 1 and 4 at 10 and 5 Gyr,
  respectively), and the combined data. Only stars with positive parallaxes
  and available radial velocities are shown. Note how the compact $E$-$L_z$
  distribution of the debris streams turns into a more complex structure in
  the lower diagrams due to the way the astrometric and radial velocity errors
  propagate. The parallax errors (see Fig.~\ref{fig:parerr}) are by far the
  dominant contribution to these structures.}
\label{fig:elz}
\end{figure*}

Various methods have been proposed and used to recover satellite remnants from
surveys of Galactic phase space. Figure~\ref{fig:elz} shows an example of
energy vs angular momentum ($E$-$L_z$) diagrams. For this figure we combined
18 million stars from the Monte Carlo model of the Galaxy with the satellite
models 1 (at 10~Gyr) and 4 (at 5~Gyr). The value of $M_{V,\mathrm{tracer}}$ is
$0.4$ and $1.0$ respectively. Keep in mind that the figures under-represent
the real contrast between Galaxy and satellite.

\section{Future work}

We intend to use this work to study in detail the retrieval of debris streams
from the Gaia catalogue by making use of the accurate phase space data that
Gaia will provide throughout the Galaxy. However, as can be appreciated from
the $E$-$L_z$ diagrams shown in Fig.~\ref{fig:elz} it will not be possible to
trace complete debris streams based on phase space data alone. Fortunately,
using the photometric information from Gaia will provide astrophysical
parameters for the different stellar populations. This will enable a drastic
narrowing down of the parts of the Gaia catalogue that need to be
searched. Investigating this will require improvements to the simulations.

The simulation of actual Gaia photometry from the broad and medium band
photometers is possible using the tools developed at the University of
Barcelona (see the contributions by Jordi et~al.\ and Carrasco et~al.\ in this
volume). An extinction model can be included in an approximate way using
currently available 3D extinction models such as the one presented by Drimmel
et~al.\ (this volume). Finally, if in reality the Galactic halo has not had
time to relax and wipe out the clumpy remnants of past merger events, it will
be necessary to replace the smooth halo component in our Milky Way model with
a clumpy one.

\section*{Acknowledgements}
A.B.\ thanks everyone at IA-UNAM in Ensenada for their hospitality during two
visits in which most of the work described here was done. H.V.\ and L.A.\
acknowledge support from DGAPA/UNAM grant IN113403.
\end{document}